\documentstyle[12pt]{article}

\newcommand{\beq}{\begin{equation}}
\newcommand{\eeq}{\end{equation}}
\newcommand{\bea}{\begin{eqnarray}}
\newcommand{\eea}{\end{eqnarray}}

\newcommand{\gsim}{\lower.7ex\hbox{$
\;\stackrel{\textstyle>}{\sim}\;$}}
\newcommand{\lsim}{\lower.7ex\hbox{$
\;\stackrel{\textstyle<}{\sim}\;$}}

\newcommand{\La}{\overline{\Lambda}}

\newcommand{\eod}{\end{document}}

\begin{document}
\thispagestyle{empty}
\vspace*{-20mm}

\begin{flushright}
UND-HEP-07-BIG\hspace*{.08em}05\\
DESY 07-116\\
SFB/CPP-07-45\\
LPT-ORSAY 07-67\\
LAL 07-109\\

\end{flushright}
\vspace*{7mm}

\begin{center}
{\LARGE{\bf
Memorino on the `1/2 vs. 3/2 Puzzle' in 
$\bar B \to l \bar \nu X_c$ -- a Year Later and a Bit Wiser}}
\vspace*{14mm}

{\large{\bf I.I.~Bigi$^{\,a}$, B. Blossier$^b$, A. Le Yaouanc$^c$, 
L. Oliver$^c$, O. P\`ene$^c$, J.-C. Raynal$^c$,  
A. Oyanguren$^d$, P. Roudeau$^d$}}\\
\vspace{4mm}

$^a$ {\sl Department of Physics, University of Notre Dame du Lac}
\vspace*{-.8mm}\\
{\sl Notre Dame, IN 46556, USA}\vspace*{1.5mm}\\ 
$^b$ {\sl DESY, Platanenallee 6, D-15738 Zeuthen, Germany}\\
$^c$ {\sl LPTh, Univ. de Paris Sud, F-91405 Orsay CEDEX, France} \\ 
$^d$ {\sl LAL, F-91898 Orsay CEDEX, France}

\vspace*{5mm}

{\bf Abstract}\vspace*{-.9mm}\\
\end{center}

\noindent
The OPE treatment that has been so successful in describing inclusive $\bar B \to l \bar \nu X_c$ decays 
yields sum rules (in particular the Uraltsev sum rule and its higher moments)  implying the dominance of   
the $P$ wave $j_q = 3/2$ charm states in $X_c$ over their $j_q=1/2$ counterparts. This prediction is supported by other 
general arguments as well as quark model calculations, which illustrate the OPE results, and by preliminary lattice findings. Its failure would indicate a 
significant limitation in our theoretical understanding of $\bar B \to l \bar \nu X_c$. 
Some experimental issues have been clarified since a preliminary version of this note had appeared, yet the verdict on the composition of the final states  
{\em beyond} $D$, $D^*$ and the two narrow $j_q = 3/2$ resonances remains unsettled.  
Establishing which 
hadronic configurations -- $D/D^* + \pi, D/D^* + 2 \pi, ...$ -- contribute, what their quantum numbers are and their mass distributions 
will require considerable experimental effort. We explain the 
theoretical issues involved and why a better understanding of them will be of considerable value.  
Having significant contributions from a mass continuum distribution below 
2.5 GeV raises serious theoretical questions for which we have no good answer. 
Two lists are given, one 
with measurements that need to be done and one with items of theoretical homework. Some of the 
latter can be done by employing existing theoretical tools, whereas others need new ideas.  

\tableofcontents

\vspace*{10mm}
\section{Outline}

Both our theoretical and experimental knowledge on semileptonic $B$ decays have advanced 
considerably over the last 15 years. This progress can be illustrated most strikingly by the recent success in extracting the value of $|V(cb)|$ with better than 2 \% accuracy from measurements of 
inclusive $\bar B \to l \bar \nu X_c$ transitions \cite{FLAECHER,SCHWANDA,NEWBABAR}. At the same time 
some potential problem of a rather subtle nature have emerged. One concerns BR$(B \to l \nu X_c)$, which has been very well measured now 
\cite{HFAG}: 
\bea 
{\rm BR}(\bar B_d \to l^- \bar \nu X_c) &=& (10.33 \pm 0.28) \% \\
{\rm BR}(\bar B_u \to l^- \bar \nu X_c) &=& (10.99 \pm 0.28) \% 
\label{BRSLHFAG}
\eea 
While the ratio of these branching fractions is well understood in terms of the $\bar B_d - B^-$ lifetime 
difference, their absolute scale falls below early predictions 
inferred from the Heavy Quark Expansion (HQE). Yet those were based on values for the charm quark mass that appear too heavy now. Using smaller values of $m_c$ (together with a more careful definition of heavy quark masses)
and including some novel radiative corrections \cite{CZAR} enhances the rate for 
$b\to c \bar cs$ and thus lowers $BR_{SL}(B)$. It also enhances the charm content in the final state 
of $B$ decays -- in agreement with the data \cite{CHARMCONT}. A conclusive theoretical analysis of $BR_{SL}(B)$ has not been performed yet although the tools exist. Yet we do not suspect this observable to represent a real problem for theory. 

In this note we want to focus on the 
composition of the hadronic final state in semileptonic $B$ decays {\em beyond} 
$\bar B \to l \bar \nu D/D^*$. 
Understanding the nature of the 
hadronic system in the final state -- its quantum numbers as well as mass distributions -- is 
important, since well grounded theoretical expectations and predictions can and have been 
given on these issues. Heavy quark symmetry that becomes an exact symmetry of QCD in the 
limit $m_Q \to \infty$ provides at least a convenient classification scheme. 
The $S$ wave configurations $D$ and $D^*$ represent the ground states to be followed 
by four $P$ wave $[c\bar q]$ excitations. In two 
of those the light degrees of freedom carry angular momentum 
$j_q = 3/2$ resulting in narrow resonances with spin 1 and 2. For the other two one has $j_q = 1/2$ leading to broad 
resonances with spin 0 and 1. Our theoretical understanding of semileptonic $B$ decays 
tells us rather unequivocally that the $j_q = 3/2$ states should be more abundant in the final states than their $j_q = 1/2$ 
counterparts.  This prediction appears to be at variance with some data. 
We refer to this apparent conflict as the `1/2 vs. 3/2 puzzle' \cite{PUZZLE}. One also expects smallish 
contributions from other hadronic configurations. There still seems to be some tension between 
data sets concerning the mass distribution of such additional contributions and their decay patterns. 
The aim of 
this note is to explain in a concise way the arguments involved in deriving the theoretical expectations 
and the consequences of their possible failure to stimulate further 
experimental as well as theoretical studies. The issues had caught our attention and led 
to the original `Memorino' (= short memo) more than a year ago \cite{MEMO1}. While more data have 
been obtained since, and we have pondered the issues further, we find the problems now even more intriguing and in need of a resolution. The latter has to be driven by even more detailed analyses. A 
more appropriate name for this paper might now be `Memorone' (= long memo) -- alas we decided to stick with the 
original moniker.  

After giving an overview of the experimental situation for $ B \to l \nu X_c$ in Sect.\ref{EXPSIT} we marshall the theoretical arsenal for treating those decays: the operator product expansion (OPE) in Sect.{\ref{OPETR}, the BT model in 
Sect.\ref{BTMOD} and lattice QCD in Sect.\ref{LQCD} before adding other general arguments in 
Sect.\ref{GENERAL}; in Sect.\ref{EXPCOMP} we undertake a more detailed comparison of the 
theoretical predictions and expectations with the existing data on 
$\bar B \to l \bar \nu D^{(*)} + \pi$'s from ALEPH, BaBar, BELLE, CDF, DELPHI and D0;  
in Sect.\ref{NLPAT} we comment on corresponding expectations for nonleptonic $B$ decays; finally in Sect.\ref{ACTION} we list needed homework for both theorists and experimentalists. 
We aim at being as concise as reasonably possible, while providing a guide through the literature for the more committed reader.

\section{The data}
\label{EXPSIT}

About three quarters of the inclusive semileptonic $B$ width are made up by the two channels 
$\bar B \to l \nu D/D^*$, for which a recent BaBar analysis finds \cite{BABARD**}: 
\bea 
\nonumber 
\frac{\Gamma (B^- \to l^- \bar \nu D)}{\Gamma (B^- \to l^- \bar \nu D X)}
&=& 0.227 \pm 0.014 \pm 0.016, \; 
\frac{\Gamma (B^- \to l^- \bar \nu D^*)}{\Gamma (B^- \to l^- \bar \nu D X)}
= 0.582 \pm 0.018 \pm 0.030 
\label{BUDD*}
\\
\nonumber 
\frac{\Gamma (\bar B_d \to l^- \bar \nu D)}{\Gamma (\bar B_d \to l^- \bar \nu D X)}
&=& 0.215 \pm 0.016 \pm 0.013, \;  
\frac{\Gamma (\bar B_d \to l^- \bar \nu D^*)}{\Gamma (\bar B_d \to l^- \bar \nu D X)}
= 0.537 \pm 0.031 \pm 0.036 
\label{BDDD*}
\eea
(One expects $\Gamma (\bar B \to l^- \bar \nu D X)$ to saturate 
$\Gamma (\bar B_d \to l^- \bar \nu X_c)$ for all practical purposes, and this is completely consistent 
with observation.) 
This large dominance of $D$ and $D^*$ final states, the ground states of heavy quark symmetry, 
represents actually the most direct evidence that charm quarks act basically like heavy quarks in $B$ decays. This can be invoked to justify using the heavy quark 
classification already for charm and applying arguments based on the SV limit \cite{OPTICAL}. 

For the remainder we have 
\bea 
\frac{\Gamma (B^- \to l^- \bar \nu D^{**})}{\Gamma (B^- \to l^- \bar \nu D X)}
&=& 0.191 \pm 0.013 \pm 0.019
\label{BUD**}  \\
\frac{\Gamma (\bar B_d \to l^- \bar \nu D^{**})}{\Gamma (\bar B_d \to l^- \bar \nu D X)}
&=& 0.248 \pm 0.032 \pm 0.030 \; , 
\label{BDD**}
\eea
where $D^{**}$ denotes any $D^{(*)}n\pi$ combination that is {\em not} a $D^*$. The apparent sizable 
difference in the central values of these two ratios is in conflict with theoretical expectations based 
on isospin symmetry. For the latter imposes practically equal rates for the corresponding semileptonic 
channels of $B^-$ and $\bar B_d$. 
We view this difference as due to a statistical fluctuation or a systematic bias. Averaging over 
$B^-$ and $\bar B_d$ rates yields \cite{BABARD**}:  
\bea 
\nonumber 
\frac{\Gamma (\bar B \to l^- \bar \nu D)}{\Gamma (\bar B \to l^- \bar \nu D X)}
= 0.221 \pm 0.012 \pm 0.006 &,& 
\frac{\Gamma (\bar B \to l^- \bar \nu D^*)}{\Gamma (\bar B \to l^- \bar \nu D X)}
= 0.572 \pm 0.017 \pm 0.016 \\ 
\frac{\Gamma (\bar B \to l^- \bar \nu D^{**})}{\Gamma (\bar B \to l^- \bar \nu D X)}
&=& 0.197 \pm 0.013 \pm 0.013
\label{BDD*}
\eea
BELLE finds completely consistent branching ratios for 
$B^- \to l^-\bar \nu D^0/D^{*0}$ and $\bar B_d \to l^- \bar \nu D^+$. Yet their number for 
$\bar B_d \to l^- \bar \nu D^{*+}$ appears on the low side compared with the expectation based 
on isospin. As before we take that as a sign that the data are not completely mature yet rather than 
as a real effect. 

The numbers in Eq.(\ref{BDD*}) are consistent with the HFAG averages \cite{HFAG}: 
\bea 
\nonumber 
{\rm BR} (\bar B_d \to l^- \bar \nu D^+) = (2.08 \pm 0.18) \% &,& 
{\rm BR} (\bar B_d \to l^- \bar \nu D^{*+}) = (5.29 \pm 0.19) \% \\
{\rm BR} (\bar B_d \to l^- \bar \nu D^{**+}) &=& (2.8 \pm 0.3) \% \; , 
\label{HFAGAVE}
\eea
where the $D^{**}$ rate is inferred by subtracting the $D$ and $D^*$ rates from the total semileptonic 
rate. 
Another even more recent BABAR analysis \cite{NEWNEWBABAR} 
finds numbers manifestly consistent with isospin invariance: 
\bea 
{\rm BR} (B^- \to l^- \bar \nu D^0) = (2.33 \pm 0.09_{stat}\pm 0.09_{syst}) \% \\ 
{\rm BR} (B^- \to l^- \bar \nu D^{*0}) = (5.83 \pm 0.15_{stat}\pm 0.30_{syst}) \% \\
{\rm BR} (\bar B_d \to l^- \bar \nu D^+) = (2.21 \pm 0.11_{stat}\pm 0.12_{syst}) \% \\ 
{\rm BR} (\bar B_d \to l^- \bar \nu D^{*+}) = (5.49 \pm 0.16_{stat}\pm 0.25_{syst}) \%
\eea
The ratio of corresponding branching ratios is fully consistent with the $B^-$-$\bar B_d$ lifetime ratio. 

The four $P$ wave excitations $D^{3/2}_{1,2}$ and $D^{1/2}_{0,1}$ are obvious candidates 
to provide $D^{**}$ contributions. 
ALEPH \cite{ALEPHEXP} has reconstructed $D^{**}$ states decaying into 
$D^{(*)}\pi^{\pm}$. They did not observe a significant excess of events over the expected background 
in $D^{(*)+}\pi^+$ or $D^0\pi ^-$ combinations (called `wrong sign'). From the measured rate of `right sign' combinations and assuming that only $D^{**}$ decaying to $D^{(*)}\pi$ contribute (to correct for 
channels with a missing $\pi^0$) they get (with Prob$(b \to B)=(39.7 \pm 1.0)\%$): 
\beq 
{\rm BR}(\bar B \to l \bar \nu D^{**}\to l \bar \nu D^{(*)}\pi ) = (2.2 \pm 0.3 \pm 0.3) \% 
\label{ALEPHBR1}
\eeq
Assuming the $D^{3/2}_1$ state to decay only into $D^*\pi$ they find among their $D^{**}$ sample:
\bea 
{\rm BR}(\bar B \to l \bar \nu D^{3/2}_1) &=& (0.70 \pm 0.15) \% \\
{\rm BR}(\bar B \to l \bar \nu D^{3/2}_2) &<& 0.2 \% \; . 
\label{ALEPHBR2}
\eea

DELPHI has published a re-analysis of their data \cite{DELPHI} superseding their previous 
study \cite{DELPHIPREVIOUS}. Assuming only $D^{(*)}\pi$ to contribute (to correct for channels 
with a missing $\pi^0$) they obtain 
\beq 
{\rm BR}(\bar B \to l \bar \nu D^{**}\to l \bar \nu D^{(*)}\pi ) = (2.7 \pm 0.7 \pm 0.2) \% 
\label{DELPHIBR1}
\eeq
with clear evidence for two narrow states tentatively identified with $D^{3/2}$:  
\beq 
{\rm BR}(\bar B \to l \bar \nu D_1^{3/2} )= (0.56 \pm 0.10)  \% \; , \; 
{\rm BR}(\bar B \to l \bar \nu D_2^{3/2}) = (0.30 \pm 0.08)
\label{DELPHI3/2} \; , 
\eeq 

The D0 collaboration has measured production rates of narrow $D^{**}$  states in the decay 
$\bar B \to \mu ^- \bar \nu D^*\pi$. Assuming BR$(D_1^{3/2}\to D^*\pi)=100\%$ and 
BR$(D_2^{3/2}\to D^*\pi)=(30 \pm 6)\%$ they obtain \cite{DZERO}
\beq
{\rm BR}(\bar B \to \mu ^- \bar \nu D_1^{3/2}) = (0.33 \pm 0.06) \% \; , \; 
{\rm BR}(\bar B \to \mu^- \bar \nu D_2^{3/2})= (0.44 \pm 0.16) \% \;  . 
\label{D0DATA}
\eeq
These three sets of data agree in pointing to 
\beq 
{\rm BR}(\bar B \to l^- \bar \nu D_{1,2}^{3/2}) \sim 0.8 - 1 \% \; , 
\label{3/2BR}
\eeq
yet do not paint a clear picture on the relative strength of $\bar B \to l^- \bar \nu D^{3/2}_2$ vs. 
$\bar B \to l^- \bar \nu D^{3/2}_1$. 

The observed rates for $\bar B \to l^- \bar \nu D/D^*/D^{3/2}$ are in pleasing agreement with the 
theoretical predictions described below. The fact that they do not quite saturate 
$\Gamma_{SL} (\bar B)$ is not surprising. Even so one would like to learn from the data 
what these additional final states are, in particular what their distribution in mass is. If they populate the range above 2.5 GeV, then we have at least candidates for them. If however they fall mostly below 
2.5 GeV, then we can come up at best with rather exotic explanations. These are subtle questions 
concerning smallish rates. Yet answering them will teach us important and potentially very 
surprising lessons on nonperturbative dynamics. 
This will be explained 
in Sect.\ref{EXPCOMP} after having marshaled the theoretical predictions and expectations. 

\section{The OPE treatment}
\label{OPETR}

While the OPE allows to describe inclusive transitions, no 
{\em systematic} extension to exclusive modes has been given so far. Yet even so, the OPE allows to place important constraints on some exclusive rates: 
$\bar B\to l \bar \nu D^*$, $l \bar \nu D$ (the latter involving the `BPS' approximation on which 
we comment later) are the most topical and elaborated 
examples \cite{HQE,BPS}. 

OPE results can be given also for {\em sub}classes of inclusive transitions due to various  
sum rules \cite{URSR,ORSAYSR} that can genuinely be derived from QCD; hence one can infer constraints on certain exclusive contributions.   
Those can be formulated most concisely when one adopts the heavy quark symmetry classification scheme also for the charm system in the final state of semileptonic $B$ decays. 

In the limit $m_Q \to \infty$ one has heavy quark symmetry controlling the spectroscopy for mesons as follows: The heavy quark spin decouples from the dynamics, and the hadrons can be labeled by their total spin $S$ together with the angular momentum $j_q$ carried by the light degrees of freedom, namely the 
light quarks and the gluons. The pseudoscalar and vector mesons 
$D$ and $D^*$ then form the ground states of heavy quark symmetry in the charm sector with 
$j_q=1/2$. The first excited states are four P wave configurations, namely two with $j_q = 3/2$ and 
$S=2,1$ -- $D^{3/2}_2$, $D^{3/2}_1$ -- and two with $j_q = 1/2$ and $S=1,0$ -- 
$D^{1/2}_1$, $D^{1/2}_0$; the two $3/2$ states are narrow resonances and the 
two $1/2$ states wide ones.  Then there are higher states still, namely radial excitations and higher 
orbital states; furthermore there are charm final states that cannot be properly called a hadronic resonance, but are $D/D^*$ combinations with any number of pions etc. carrying any 
allowed $J^{PC}$ quantum numbers.

The $P$ wave states 
$D^{3/2}_{2,1}$ and $D^{1/2}_{1,0}$ are obvious candidates for $D^{**}$, but they should not  
saturate it completely. One actually expects QCD radiative corrections to populate the higher 
hadronic mass region above the prominent resonances through a smooth spectrum dual to a 
superposition of broad resonances. 

Up to small isospin breaking effects one predicts these semileptonic rates to be the same. 
The usual Isgur-Wise function $\xi (w)$ is the core element in describing $\bar B\to l \bar \nu D/D^*$. It can 
be generalized to describe also the production of excited charm final states in semileptonic $B$ decays: $\tau _{1/2[3/2]}^{(n)}(w_n)$ with $w_n = v_B\cdot v_{D^{(n)}}$ is the amplitude for 
$\bar B\to l \bar \nu D^{(n)}_{1/2[3/2]}$, 
where $D^{(n)}_{1/2[3/2]}$ denotes a hadronic system with open charm carrying 
$j_q = 1/2[3/2]$ and label $n$; it does {\em not} need to be a bona fide 
resonance. 

Various sum rules can be derived from QCD proper relating the moduli of these amplitudes and 
powers of the excitation energies $\epsilon ^{(n)} \equiv M_{D^{(n)}} - M_D$ to heavy quark parameters. 
Adopting the so-called `kinetic scheme', as we will throughout this note, one obtains in particular 
\cite{URSR}: 
\bea
\frac{1}{4} &=& - \sum _n \left| \tau_{1/2}^{(n)}(1)\right| ^2 + 
\sum _m \left| \tau_{3/2}^{(m)}(1)\right| ^2 
\label{SR1}
\\
\mu_{\pi}^2 (\mu)/3 &=& 
\sum _n^{\epsilon^{(n)} \leq \mu} \left( \epsilon^{(n)}_{1/2}\right) ^2\left| \tau_{1/2}^{(n)}(1)\right| ^2 +
2\sum _m^{\epsilon^{(m)} \leq \mu} \left( \epsilon^{(m)}_{3/2}\right) ^2\left| \tau_{3/2}^{(m)}(1)\right| ^2 
\label{SR2}
\\
\mu_{G}^2 (\mu)/3 &=& 
-2\sum _n^{\epsilon^{(n)} \leq \mu} \left( \epsilon^{(n)}_{1/2}\right) ^2\left| \tau_{1/2}^{(n)}(1)\right| ^2 +
2\sum _m^{\epsilon^{(m)} \leq \mu} \left( \epsilon^{(m)}_{3/2}\right) ^2\left| \tau_{3/2}^{(m)}(1)\right| ^2
\label{SR3}
\eea
where the summations go over all hadronic systems with excitation energies 
$\epsilon^{(n,m)} \leq \mu$ 
\footnote{The sum rule of Eq.(\ref{SR1}) does not require a cut-off or normalization scale 
$\mu$, as is already implied by its left hand side \cite{URSR}.}. 
The sum rules relate the observables $\mu_{\pi}^2 (\mu)$ and $\mu_G^2 (\mu)$ that 
describe the fully inclusive $\bar B \to l^- \bar \nu X_c$ transitions with sums over {\em sub}classes 
of exclusive modes. For the purpose of these sum rules it is irrelevant whether these hadronic systems 
are bona fide resonances or not; what matters is that they are $P$ wave configurations with 
$j_q = 3/2$ or $1/2$.

These sum rules allow us to make both general qualitative as well as 
(semi)quantitative statements. 
On the {\em qualitative} level we learn unequivocally that the `3/2' transitions have to dominate over the 
`1/2' ones, as can be read off from Eqs.(\ref{SR1},\ref{SR3}). 
Furthermore we know that 
\beq 
\mu_{\pi}^2 (\mu ) \geq \mu _G^2(\mu ) 
\label{INEQ}
\eeq  
has to hold for any $\mu$ 
\cite{OPTICAL}, as is obvious from Eqs.(\ref{SR2},\ref{SR3}).  
On 
the {\em quantitative} level  it is not a priori clear, at which scale 
$\mu$ these sum rules are saturated and by which kind of states. We will address these issues below. 

We have learnt a lot about the numerical values of the heavy quark parameters: the most accurate value for the chromomagnetic moment $\mu_G^2$ can be deduced from the $B^*-B$ hyperfine mass splitting: 
\beq 
\mu_G^2(1 \; {\rm GeV}) = (0.35 \pm 0.03) \; {\rm GeV}^2
\label{HYPER}
\eeq

The analyses of Refs.\cite{FLAECHER,SCHWANDA,NEWBABAR} based on 
comprehensive study of energy and hadronic mass moments in $\bar B \to l \bar \nu X_c$ yield the 
following values: 
\bea 
\mu_G^2(1 \; {\rm GeV}) &=& 
\left\{
\begin{array}{ll}
0.297 \pm 0.024|_{exp} \pm 0.046|_{HQE} \; {\rm GeV}^2 &\cite{FLAECHER} \\
0.358 \pm 0.060|_{fit} \pm 0.003|_{\delta \alpha_S} \; {\rm GeV}^2 & \cite{SCHWANDA}\\
0.330 \pm 0.042|_{exp} \pm 0.043|_{theo}  \; {\rm GeV}^2 &\cite{NEWBABAR}
\end{array}
\right. \\
\mu_{\pi}^2(1 \; {\rm GeV}) &=& 
\left\{
\begin{array}{ll}
0.401 \pm 0.019|_{exp} \pm 0.035|_{HQE} \; {\rm GeV}^2 &\cite{FLAECHER} \\
0.557 \pm 0.091|_{fit} \pm 0.013|_{\delta \alpha_S} \; {\rm GeV}^2 & \cite{SCHWANDA}\\
0.471 \pm 0.034|_{exp} \pm 0.062|_{theo} \; {\rm GeV}^2 &\cite{NEWBABAR}
\end{array}
\right. 
\eea
These experimental numbers are consistent with each other and with the predictions of 
Eqs.(\ref{HYPER},\ref{INEQ}) with{\em out} the latter having been imposed. All three determinations of  
$\mu_{\pi}^2 (1\; {\rm GeV})$ are within one sigma of 0.45 GeV$^2$, which we will use as a reference point for our subsequent considerations:  
\beq 
\mu_{\pi}^2 (1\; {\rm GeV})|_{\rm ref} = 0.45 \; {\rm GeV}^2 \; . 
\label{REFP}
\eeq

For the lowest excitation energies we have 
\beq 
\epsilon^{(0)}_{3/2} \sim 450 \; {\rm MeV} \; \; , \; \; \epsilon ^{(0)}_{1/2} \sim (300 - 500) \; {\rm MeV} \; ; 
\label{EPS}
\eeq
these values for $\epsilon ^{(0)}_{1/2}$ allow for `1/2' states to be both lighter and heavier than the 
narrow `3/2' states. 

From Eqs.(\ref{SR2},\ref{SR3}) one obtains 
\bea 
\mu_{G}^2 (\mu) &=& 6  \sum _n^{\epsilon^{(n)} \leq \mu} \left( \epsilon_{3/2}^{(n)}\right) ^2\left| \tau_{3/2}^{(n)}(1)\right| ^2 
- \frac{2}{3} \left( \mu_{\pi}^2 (\mu) - \mu_{G}^2 (\mu) \right) 
\label{SRMUG}\\
\mu_{\pi}^2 (\mu) - \mu_{G}^2 (\mu) &= &
9 \sum _n^{\epsilon^{(n)} \leq \mu} \left( \epsilon_{1/2}^{(n)}\right)^2\left| \tau_{1/2}^{(n)}(1)\right| ^2
\label{SRDIFF}
\eea
as convenient expressions to read off natural `scenarios' for the implementation of the OPE description and its sum rules. 

One can reasonably assume these sum rules to be saturated {\em approximately} by the lowest states 
$n=0$ for $\mu \leq 1$ GeV. 
This rule of thumb (not to be confused with sum rules) is based on general experience with sum rules and on considerations of how 
$\mu_{\pi}^2 (\mu)$ and $\mu_{G}^2 (\mu)$ vary with the scale $\mu$. It does not mean that the various sum rules would saturate to the same degree at a given $\mu$; for they reflect different dynamical situations.

Using Eq.(\ref{REFP}) we then infer from Eq.(\ref{SRMUG}) 
\beq 
\tau_{3/2}^{(0)}(1) \sim 0.6 
\label{3/2EST}
\eeq
and from Eq.(\ref{SRDIFF}) 
\beq  
\tau_{1/2}^{(0)}(1) \leq 0.14 - 0.32 \; , 
\label{1/2EST}
\eeq
which are reasonable numbers as our subsequent considerations will illustrate. 
For $\mu_{\pi}^2 (1\; {\rm GeV}) = 0.4$ GeV$^2$ one 
has $\tau_{1/2}^{(0)}(1) \leq 0.1 - 0.2$ and 
$ \tau_{3/2}^{(0)}(1) \sim 0.56$. On the other hand  
$\mu_{\pi}^2 (1\; {\rm GeV}) = 0.55$ allows for a sizeable production rate of the lowest `1/2' state -- 
$\tau_{1/2}^{(0)}(1) \leq 0.2 - 0.45$ -- although does not enforce it, and 
$\tau_{3/2}^{(0)}(1) \sim 0.63$. These numbers for $ \tau_{3/2}^{(0)}(1)$ should also be seen more like 
an upper bound, since the assumed saturation of the sum rules can hardly be exact. 

Imposing also 
Uraltsev's sum rule, Eq.(\ref{SR1}), with (approximate) saturation assumed 
for the lowest states  
$\left( \left| \tau_{1/2}^{(0)}(1)\right| ^2 \simeq \left| \tau_{3/2}^{(0)}(1)\right| ^2 -  \frac{1}{4}\right)$ leads to 
\beq 
\tau_{3/2}^{(0)}(1)|_{\rm SR} \sim 0.6 \; , \; \tau_{1/2}^{(0)}(1)|_{\rm SR} \sim 0.32 \; ; 
\label{OPESREST}
\eeq
i.e., very much the upper end of Eq. (\ref{1/2EST}) inferred from 
$\mu_{\pi}^2 (1\; {\rm GeV}) = 0.45$ GeV$^2$ and pointing to 
$\epsilon ^{(0)}_{1/2} \sim 300$ MeV, i.e. 
a relatively low mass for the `1/2' states. A scenario with a low 
$\mu_{\pi}^2 (1\; {\rm GeV}) = 0.4$ GeV$^2$ 
is however hardly compatible with it, whereas a high $\mu_{\pi}^2 (1\; {\rm GeV}) = 0.55$ GeV$^2$ 
can be accommodated with $\epsilon ^{(0)}_{1/2} \sim 450$ MeV. 

While these numbers are reasonable, one cannot rule out significant 
contributions  from higher states like the first {\em radial} excitations of the $P$ wave states. 
We will address such a scenario in Sect. \ref{BTMOD}.

In summary: 
\begin{itemize}
\item 
The OPE treatment leads to the general prediction that among $P$ wave configurations production of `3/2' states dominates over that of `1/2' states in semileptonic $B$ decays and that the former yields a 
significant contribution. 
\item 
Approximate saturation of the sum rules by the $P$ wave states represents a scenario consistent with the data and leads to   
semi-quantitative estimates for the degree of the dominance of `3/2' over `1/2' production. 
\item 
At the same time there is no reason from the OPE to expect that $D$, $D^*$ and the two narrow $D_{3/2}$ states 
saturate the semileptonic width. We know that the OPE treatment describes the hadronic mass moments 
in semileptonic $B$ decays very sucessfully \cite{FLAECHER,SCHWANDA,NEWBABAR}, and the data clearly show 
BR$(\bar B \to l^- \bar \nu D_X) \sim 1 - 2 \%$ with 
$D_X \neq (D, D^*, D^{(3/2)}_{1,2})$.  Already on general grounds one would not expect $D_X$ to be mostly a narrow resonance. More specifically one can start from the numbers in 
Eq. (\ref{HFAGAVE}) and split the $\bar B \to l^- \bar \nu D^{**}$ contribution into two components 
with BR$(\bar B \to l^- \bar \nu D^{(3/2)}) = 0.8\%$ and 
BR$(\bar B \to l^- \bar \nu D_X) = 2\%$. From the hadronic mass moments determined in the 
kinetic scheme \cite{FLAECHER,SCHWANDA} one can then infer by matching what the hadronic 
mass moments for the $D_X$ contribution have to be. One typically finds 
$\langle M(D_X)\rangle \sim 2.4 - 2.6$ GeV with a spread of about 200 MeV. DELPHI has inferred the mass moments of the $D^{**}$ contributions and found a central value of 2.5 GeV with a spread of  
230 MeV. 

Then the question arises what makes up this $D_X$ contribution. Some broad $`3/2'$ configuration
presumably of a non-renonant nature? Or states that do {\em not} contribute to the sum rules like 
$J^P =0^-,1^-$ states? Radial excitations would fit this bill, yet run counter to arguments to be discussed  in Sect. \ref{EXPCOMP}. 

\item 
No reliable prediction on their decay patterns -- i.e. whether they yield $D^{(*)} \pi$ or 
$D^{(*)} 2\pi$ etc. -- can be inferred from 
the OPE treatment per se. 
\end{itemize} 

\section{The BT model}
\label{BTMOD}

Based on the OPE treatment alone one cannot be more specific numerically. 
To go further one relies on quark models for guidance. The dominance of the `3/2' over the `1/2' states emerges naturally in 
all quark models obeying known constraints from QCD as well as Lorentz covariance. 
This can be demonstrated  explicitly with the Bakamjian-Thomas covariant quark model  
\cite{BTMOD}, which satisfies heavy quark symmetry and the Bjorken as well as spin sum rules referred to above. It allows 
to determine the masses of various charm excitations and 
to compute the production rates in semileptonic  \cite{BTORSAY1,BTORSAY2,HYCHENG} 
as well as nonleptonic $B$ decays \cite{BTORSAY2}. 
The BT model provides a quantitative illustration of the heavy quark limit, in particular concerning 
the sum rule of Eq.(\ref{SR1}). One finds 
\bea 
 \tau ^{(0)}_{1/2}(1)|_{\rm BT} &=& 0.22  \\
\tau ^{(0)}_{3/2}(1)|_{\rm BT} &=& 0.54  
\label{BTEST}
\eea
together with predictions for the slopes. These values are fully consistent with the estimates 
given above. 
For the semileptonic modes the BT model yields: 
\bea 
{\rm BR}(\bar B \to l \bar \nu D) &=& (1.95 \pm 0.10) \% \; , \; 
\label{BRTH1}
\\
{\rm BR}(\bar B \to l \bar \nu D^*) &=& (5.90 \pm 0.20) \% 
\label{BRTH2}
\\
{\rm BR}(\bar B \to l \bar \nu D_2^{3/2}) &=& (0.63^{+0.3}_{-0.2}) \% 
\label{BRTH3} 
\\
{\rm BR}(\bar B \to l \bar \nu D_1^{3/2} )&=& (0.40^{+0.12}_{-0.14}) \%
\label{BRTH4}
\\
{\rm BR}(\bar B \to l \bar \nu D_1^{1/2}) &=& (0.06 \pm 0.02) \% 
\label{BRTH5} 
\\
{\rm BR}(\bar B \to l \bar \nu D_0^{1/2}) &=& (0.06 \pm 0.02) \%
\label{BRTH6}
\eea

The following basis and features of these predictions should be noted:  
\begin{itemize}
\item 
The predictions for $\bar B\to l \bar \nu D/D^*$ are based on 
the following parametrization of the 
Isgur-Wise function: $\xi (w) = \left( \frac{2}{w+1}\right) ^{2\rho_2}$, where 
$\rho_2$ denotes its slope. For the latter we have used the value from   
Ref. \cite{BABSLOPE}. The predictions agree with the data. 
\item 
The branching ratios for the $P$ wave states and the theoretical uncertainties are obtained 
by using the BT model value for the form factors for $w=1$ and allowing 
for a $\pm 50$ \% variation in the slope given by the BT model. This is the origin of the large 
relative errors in the predicted rates. 
We find a strong dominance of `3/2' over `1/2' production for the $P$ wave states 
as inferred already from 
the Sum Rules.     
\item 
These branching ratios add up  to $9.00 \pm 0.40$ \% and thus fall  short of saturating the observed 
$\Gamma_{SL}(B)$, Eq. (\ref{BRSLHFAG}).  
Such a deficit is not surprising as argued before. The more specific question is whether the BT model can account for the required additional width. The answer is not known yet. What can be said is that the hadronic mass for these extra contributions cannot be much lower than 2.6 GeV and that a sizable number of channels might be involved.  
\item 
The model as it is does not allow to compute $1/m_Q$ corrections; i.e., effectively it treats the 
$m_Q \to \infty$ limit, since only then it is covariant.

\end{itemize}
With an explicit quark model one can address also higher states. The BT model finds for the first 
{\em radial} excitations of the `3/2' and `1/2' $P$ wave states still sizable amplitudes 
$\tau _{3/2}^{(1)} \simeq 0.21$ and $\tau _{1/2}^{(1)} \simeq 0.20$. 
While these states enhance $\mu_{\pi}^2( 1 \; {\rm GeV})$ significantly, their 
contributions to $\Gamma_{SL}(B)$ are still found to be insignificant.

\section{Lattice QCD}
\label{LQCD}
In principle the two form factors $\tau_{1/2}(1)$ and $\tau_{3/2}(1)$ can be computed in a straightforward way using the HQET equation of motion $(v\cdot D) \,h_v=0$ \cite{lig}:
\bea\nonumber
_v\langle 0^+| \bar{h}_v \gamma^i \gamma^5 D^j h_v|0^-\rangle_v &=& i\, g^{ij}\,
\tau_{1/2}(1)\, (\La_{0^+}-\La_{0^-}),\\
_v\langle 2^+| \bar{h}_v \left(\frac{\gamma^i \gamma^5 D^j+\gamma^j \gamma^5 D^i}{2}\right) 
h_v|0^-\rangle_v &=& -i\sqrt{3}\,\epsilon^{*ij}\, \tau_{3/2}(1)\, (\La_{2^+}-\La_{0^-}),
\eea
where $v=(1,\vec{0})$ is the heavy quark velocity, $\epsilon^*$ the 
polarization tensor of the $2^+$ state and
$\La_{\rm J^P}$ the dominant term in the OPE expression for the ${\rm J^P}$ heavy-light meson  
binding energy. On 
the lattice the covariant derivative $D_i$ applied to the static quark field $h(\vec{x},t)$ is 
expressed as 
$D_i\, h (\vec x,t)\to \frac 1{2a}\left (U_i(\vec x,t)h (\vec x+\hat i,t) 
- U_i^\dagger(\vec x-\hat i,t) h (\vec x-\hat i,t) \right )$; $U_i(\vec x,t)$ denotes the gauge link. 
One calculates as usual the two-point functions $C^2_{\rm J^P}(t) =
\langle 0|O_{\rm J^P}(t)O_{\rm J^P}^\dag(0)|0\rangle$,  the three-point functions 
$C^3_{\rm J^P,0^-}(t_1,t_2)=\langle 0|O_{\rm J^P}(t_2)O_\Gamma(t_1)O_{0^-}^\dag(0)|0\rangle$ 
and  
$\langle J^P|O_\Gamma|0^-\rangle \propto 
R(t_1,t_2)=\frac{C^3_{\rm J^P,0^-}(t_1,t_2)}{C^2_{0^-}(t_1)C^2_{\rm J^P}(t_2-t_1)}$.

Alas, numerical complications appear, because orbital as well as radial excitations can contribute. 
To extract properly the matrix element for the P wave state 
$\langle J^P|O_\Gamma|0^-\rangle$, one has to effectively suppress 
the coupling of radial excitations (with quantum numbers $n>1,\, J^P$) 
to the vacuum. This can be achieved by choosing an appropriate interpolating field 
$O_{\rm J^P}$ such that $\langle n>1\, J^P|O_{\rm J^P}|0 \rangle=0$ holds or by having 
huge statistics to diminish statistical fluctuations at large times (where 
the fundamental state is no more contaminated by radial excitations). 
This poses a problem in particular for the $2^+$ state, for which the usual interpolating 
field seems to couple also the  first radial excitation quite strongly to the vacuum. 
Moreover reaching the required stability of $R(t_1,t_2)$ as a function of $t_2$ poses a 
serious challenge. Hopefully all to all propagators technology will be of great help, as it has
already proved to be in studies of the static-light spectrum \cite{UKQCD-Trinlat} and in the 
determination of hadronic matrix elements \cite{Onogi}.

We will need very careful and dedicated lattice studies to obtain meaningful and reliable 
results for $\tau_{3/2,1/2}$. As an already highly relevant intermediate step 
one can concentrate first on $\tau_{1/2}$ to see whether lattice QCD confirms its suppression 
as inferred from both the sum rules and the BT model. A preliminary study in
the quenched approximation with $\beta=6.0\, (a^{-1}=2 \,{\rm GeV^{-1}})$ and
$m_q \simeq m_s$ yields  
\cite{tauorsay}: 
\bea 
\tau_{1/2}^{(0)}(1)|_{\rm LQCD} &\sim& 0.41 \pm 0.05 \\
\tau_{3/2}^{(0)}(1)|_{\rm LQCD} &\sim& 0.57 \pm 0.10 \; ,  
\eea
where only the statistical errors are given.  Again we find dominance of the `3/2' over the `1/2' 
amplitude even in numerical agreement with the values inferred from the sum rules, 
see Eq.(\ref{OPESREST}); $\tau_{1/2}^{(0)}(1)|_{\rm LQCD}$ appears significantly larger than 
the BT estimate in Eq.(\ref{BTEST}). 

Apart from unquenching and lowering the value of $m_q$  one can improve and
refine this analysis also by simulating a {\em non-static} charm quark,  i.e.
applying HQET to the $B$ meson only. This would allow to evaluate $1/m_c$
corrections. However the first improvement has to be done with considerable care: 
While unquenched simulations with light quark masses lower 
than $m_s/5$ (i.e. a corresponding pseudoscalar Goldstone boson mass lower than $300$ MeV) 
have become customary now, one has to worry about a possible mixing between a $D^{**}$ resonance and a $D \pi$ state.

\section{Two other general arguments on $|\tau_{1/2}/\tau_{3/2}|^2$}
\label{GENERAL}

The numerics of the theoretical predictions on semileptonic $B$ decays given above have to be taken `cum grano salis'. Yet their principal feature -- the preponderance of `3/2' over `1/2' states -- has to be taken very seriously, since they are a general consequence of the OPE treatment. 
It is further supported  by two rather general observations that point in the same direction as the detailed theoretical considerations given above:  
\begin{itemize}
\item 
When interpreting data one should keep in mind that the contributions of $D^{1/2}_{1,0}$ to  
$\Gamma (\bar B \to l \bar \nu D^{**})$ 
are suppressed relative to those from  $D^{3/2}_{2,1}$ by a factor of two to three 
due to kinematics \cite{BTORSAY1}. 
Thus one finds for reasonable values of $\tau^{(0)}_{1/2}$ that 
$\Gamma(\bar B \to l \bar \nu D^{1/2})$ 
falls below $\Gamma (\bar B \to l \bar \nu D^{3/2})$ by one order of magnitude, as illustrated above, see 
Eqs.(\ref{BRTH5},\ref{BRTH6}). 
For the two widths to become comparable, one would need a greatly enhanced 
$\tau^{(0)}_{1/2}$. 
\item 
There is a whole body of evidence showing that in so-called class I nonleptonic $B$ decays like 
$\bar B_d \to D^{(*)+}\pi^-$ naive factorization provides a very decent description of the data. Invoking 
this ansatz also for $\bar B_d \to D^{**+}\pi^-\to D^{(*)0} \pi ^+\pi^-$ one infers from BELLE's data 
\cite{BELLENL} that the production of `1/2' states appears to be strongly suppressed relative to that 
for `3/2' ones. It implies that  $|\tau_{1/2}/\tau_{3/2}|^2$ is small and certainly less than unity. 
The same feature is found in more recent measurements from BABAR \cite{BABARNL} 
\footnote{The comparison of the theoretical predictions with the measured rates is not 
straightforward, since the data are given as products of production and decays branching ratios, and one has to use the model to calculate both. The BABAR analysis also lumps together the decays of more than one state.}. 
This agrees with the theoretical expectations described before; more importantly it shows 
in a rather model independent way that there is no large unexpected enhancement of  
$|\tau_{1/2}|$. Those values also allow to saturate the sum rule of Eq.(\ref{SR1}) 
within errors already with the $n=0$ states. 

The form factors are actually probed at $w=1.3$ in this nonleptonic transition; yet a natural functional  dependence on $w$ supports this conclusion to hold for $1 \leq w \leq 1.3$ in semileptonic 
channels. 
\end{itemize}
These arguments are based on the heavy quark mass limit. The as yet unknown finite mass corrections 
could modify these conclusions somewhat. 

\section{Detailed comparison with the data on semileptonic $B$ decays}
\label{EXPCOMP}

Different experiments and theoretical 
treatments agree on 
\begin{itemize}
\item
$\Gamma (\bar B \to l^- \bar \nu X_c)$ being dominated by the two modes 
$\bar B \to l^-\bar \nu D/D^*$,  
\item 
with $D^{(3/2)}$ final states providing about 10\% to it and
\item
BR$(\bar B \to l^-\bar \nu X_c) \simeq (1 - 2)\%$ has to come from other hadronic 
configurations $D_X$. 

\end{itemize}
The question  still open concerns the nature of this last component $D_X$. 

At first sight one might wonder why one should worry about the identification of channels that sum up to no more than 2\% in overall branching ratio. Yet they constitute 10 - 20\% of all semileptonic transitions, and 
-- maybe more importantly -- theory makes quite non-trivial statements about them. We can learn 
important lessons about non-perturbative dynamics, even if those predictions are refuted by 
experiment.

Theory makes a rather robust 
prediction that it can{\em not} come from the broad $P$ wave states $D^{(1/2)}$. The OPE framework by itself 
can accommodate all three features listed above, as long as $D^{(1/2)}$ is insignificant in 
the third item. It points to hadronic contributions that are broad in mass with{\em out} a firm prediction 
on their average mass -- both $\langle M(D_X)\rangle \leq 2.4$ GeV or $>2.5$ GeV 
seem {\em a priori} feasible -- or their decay patterns; i.e. $D^{(*)}\pi$ vs. $D^{(*)}\pi\pi$ 
(vs. $D^{(*)}\eta$ etc.). 

Both ALEPH and DELPHI can account for all of $\Gamma (\bar B \to l^- \bar \nu X_c)$ 
with $\bar B \to l^-\bar \nu D/D^*$ and $\bar B \to l^- \bar \nu D^{**} \to l^- \bar \nu D^{(*)}\pi$, 
see Eqs.(\ref{ALEPHBR1},\ref{DELPHIBR1}) with no established signal for $D^{(*)}\pi \pi$ states 
contributing. ALEPH places relatively tight bounds on higher 
combinations from the observed number of `wrong sign' combinations: 
\beq 
{\rm BR}(\bar B \to l \bar \nu D^{*}\pi \pi) \leq 0.35 \% \; , \; {\rm BR}(\bar B \to l \bar \nu D\pi \pi) \leq 0.9 \%
 \; \; \; ({\rm 90 \% \, C.L.}) \, ; 
\label{ALEPHBR3}
\eeq
DELPHI's bounds are less tight: 
\beq 
{\rm BR}(\bar B \to l \bar \nu D^{*}\pi \pi) \leq 1.2 \% \; , \; {\rm BR}(\bar B \to l \bar \nu D\pi \pi) \leq 1.3 \%
\label{DELPHIBRPIPI}
\eeq
Considering that $1^+$ $D^{**}$ can decay into $D\pi \pi$ and analyzing the $D\pi$ mass 
distribution DELPHI fits a value of $(19\pm 13)\%$ for this component. In their analysis of hadronic 
mass moments such a possibility has been included with 
BR$(\bar B \to l \bar \nu D\pi\pi) = (0.36 \pm 0.27)\%$. This turns out to be the dominant systematic uncertainty 
in their hadronic mass moment measurement.

DELPHI found a significant rate for producing a broad hadronic mass distribution in $D^{(*)}\pi$:  
\beq 
{\rm BR}(\bar B \to l \bar \nu D_{"1"}) = (1.24 \pm 0.25 \pm 0.27) \% \; , \; 
{\rm BR}(\bar B \to l \bar \nu D_{"0"})= (0.65 \pm 0.69) \% \;  .   
\label{DELPHI1/2}
\eeq
{\em If} the broad contributions were indeed to be identified with the $D_{1,0}^{1/2}$ as 
already implied in Eq.(\ref{DELPHI1/2}) -- an a priori reasonable working 
hypothesis -- one would have a clear cut and significant conflict with the OPE expectations 
as well as the numerically more specific BT model predictions, see Eqs.(\ref{BRTH5}, \ref{BRTH6}). For DELPHI's data would yield  
$\Gamma (\bar B \to l \bar \nu D^{1/2}) > \Gamma (\bar B \to l \bar \nu D^{3/2})$. This conflict has been referred to 
as the `1/2 $>$ 3/2 puzzle' \cite{PUZZLE}. Since, as sketched before, the theoretical predictions are based on a 
rather solid foundation, they should not be discarded easily.  Of course there is no proof that the broad $D/D^*+\pi$ systems are indeed the 
$j_q = 1/2$ P wave states; they could be radial excitations or non-resonant combinations of undetermined quantum numbers.   Thus the DELPHI data taken by themselves are 
{\em not necessarily} in 
conflict with theoretical expectations.

However the plot thickens in several experimental as well as theoretical respects: 
\begin{itemize}
\item 
In 2005 BELLE has presented an analysis of $\bar B\to l \bar \nu D/D^* \pi$ \cite{BELLESL}, which appears to be in conflict with previous findings. Reconstructing one $B$ completely in 
$\Upsilon (4S) \to B \bar B$, they analyze the decays of the other beauty meson and obtain: 
\bea 
{\rm BR}(B^- \to l^- \bar \nu D\pi ) &=& (0.81 \pm 0.18) \% 
\label{BELLESL1}\\
{\rm BR}(B^- \to l^-  \bar \nu D^*\pi ) &=& (1.00 \pm 0.22) \% 
\label{BELLESL2}\\
{\rm BR}(\bar B_d \to l^-  \bar \nu D\pi ) &=& (0.49 \pm 0.13) \% 
\label{BELLESL3}\\
{\rm BR}(\bar B_d \to l^- \bar \nu D^*\pi ) &=& (0.97 \pm 0.22) \%
\label{BELLESL4}
\eea
BELLE's separation of final states with $D$ and $D^*$ is of significant value, since it 
provides an indirect and model dependent handle on $`3/2'$ and $`1/2'$ production. 
For with the help of a quark model one can calculate both the production rates for the 
$D_{1,2}^{(3/2)}$ and $D_{0,1}^{(1/2)}$ and their branching fractions into $D\pi$ and $D^*\pi$. 
BELLE's numbers are actually quite consistent with the theoretical predictions the BT model yields for `3/2' P wave production. It is of course still desirable for BELLE to determine the quantum numbers of their  hadronic final states.

Combining the two classes of final states they arrive at: 
\bea 
{\rm BR}(B^- \to l^- \bar \nu D^{(*)}\pi ) &=& (1.81 \pm 0.20 \pm 0.20) \% \\
{\rm BR}(\bar B_d \to l^- \bar \nu D^{(*)}\pi ) &=& (1.47 \pm 0.20 \pm 0.17) \%
\eea
leaving room for a large $D^{(*)}\pi\pi$ component of $\sim (1.3 \pm 0.4)\%$, whereas 
previous studies have obtained 90\% C.L. upper limits ranging from $0.35$ to $1.3$ \%, 
see Eqs.(\ref{ALEPHBR3},\ref{DELPHIBRPIPI}). 
\item 
BABAR's most recent analysis \cite{NEWNEWBABAR} yields rather consistent numbers: 
\bea
{\rm BR} (B^- \to l^- \bar \nu D\pi) &=& (0.63 \pm 0.09_{stat}\pm 0.05_{syst}) \% \\ 
{\rm BR} (B^- \to l^- \bar \nu D^*\pi) &=& (0.89 \pm 0.08_{stat}\pm 0.06_{syst}) \% \\
{\rm BR} (\bar B_d \to l^- \bar \nu D\pi) &=& (0.65 \pm 0.12_{stat}\pm 0.05_{syst}) \% \\ 
{\rm BR} (\bar B_d \to l^- \bar \nu D^*\pi) &=& (0.72 \pm 0.12_{stat}\pm 0.06_{syst}) \% \; , 
\eea
which can be combined to 
\bea 
{\rm BR}(B^- \to l^- \bar \nu D^{(*)}\pi ) &=& (1.52 \pm 0.12 \pm 0.10) \% \\
{\rm BR}(\bar B_d \to l^- \bar \nu D^{(*)}\pi ) &=& (1.37 \pm 0.17 \pm 0.10) \%
\eea
again leaving room for a significant  $D^{(*)}\pi\pi$ component of about 1.3\%. 
\item 
Using DELPHI's numbers stated in Eqs.(\ref{DELPHI3/2}, \ref{DELPHI1/2}) and assuming 
that the "1" and "0" state decay 100 \% into $D^*\pi$ and $D\pi$, respectively, one arrives at 
\bea 
{\rm BR}(B^- \to l^- \bar \nu D\pi ) &=& {\rm BR}(\bar B_d \to l^- \bar \nu D\pi ) \sim  
(0.9 \pm 0.7)  \% \\
{\rm BR}(B^- \to l^- \bar \nu D^*\pi ) &=& {\rm BR}(\bar B_d \to l^-  \bar \nu D^*\pi ) \sim 
(1.9 \pm 0.4) \%  \; . 
\eea
for a total of 
\beq 
{\rm BR}(B^- \to l^- \bar \nu D^{(*)}\pi ) = {\rm BR}(\bar B_d \to l^-  \bar \nu D^{(*)}\pi ) \sim 
2.8 \%
\eeq
One should note that the qualitative trend is the same as with BELLE's findings, 
Eqs.(\ref{BELLESL1} - \ref{BELLESL4}) -- namely that $D^*\pi$ final states dominate over 
$D\pi$ ones -- yet the total $D^{(*)}\pi$ rate exceeds that reported by BELLE.

\item 
The BT model predicts for $D\pi$ and $D^*\pi$ production: 
\bea 
{\rm BR}(B^- \to l^- \bar \nu D\pi ) &=& {\rm BR}(\bar B_d \to l^- \bar \nu D\pi ) = 
0.51  \% \\
{\rm BR}(B^- \to l^- \bar \nu D^*\pi ) &=& {\rm BR}(\bar B_d \to l^-  \bar \nu D^*\pi )=
0.65 \% \; , 
\eea
which is on the low side of BELLE's numbers and a forteriori for DELPHI's findings, but consistent 
with BABAR's data. This is of course a rephrasing of the `1/2 vs. 3/2' puzzle.

\item 
In the BPS approximation \cite{BPS} one has $\tau ^{(n)}_{1/2}=0$. Assuming that the 
sum rule of Eq.(\ref{SR1}) saturates already with the $n=0$ state, one obtains 
$\tau ^{(0)}_{3/2}=\frac{1}{2}$ leading to 
\bea 
{\rm BR}(B^- \to l^- \bar \nu D\pi ) &=& {\rm BR}(\bar B_d \to l^- \bar \nu D\pi ) = 
0.39  \% \\
{\rm BR}(B^- \to l^- \bar \nu D^*\pi ) &=& {\rm BR}(\bar B_d \to l^-  \bar \nu D^*\pi )=
0.50 \%  \; ; 
\eea
i.e., lower still. This might not be that surprising, since the BPS ansatz is at best an approximation 
rather than a systematic expansion.

\item 
So far there is no experimental evidence for high mass hadronic states. 
One finds  
that about 6.4\% and 18.3\% of all $D^{**}$ states have masses between 2.6 and 3.3 GeV for 
the CDF and DELPHI data, respectively, which drop to 3.2 \% and 7.8\% for the mass range 
2.8 to 3.3 GeV and 0.3 \% and 3.1 \% for 3.0 to 3.3 GeV. 

On the other hand, CDF seems to see more events below the $D^{3/2}$ peaks.  
Such low mass $D^{(*)}\pi$ events could be due to higher mass states decaying into 
$D^{(*)}\pi\pi$. CDF has not incorporated this scenario into their analysis, since previous measurements showed no evidence for such decays. 

\item 
One would conjecture that if the observed mass spectrum indeed differs significantly from theoretical expectations -- in its center of gravity as well as its spread --, then the measured hadronic mass moments should not follow theoretical predictions -- 
yet they do 
\cite{FLAECHER,SCHWANDA,NEWBABAR,DELPHI,BABARVCB,BATTAG,CDFMOM}. 

\item 
There is a more general problem. We have said before that the OPE treatment could {\em a priori}  accommodate significant contributions with hadronic systems exhibiting a rather broad distribution in mass and centered below 2.5 or even 2.4 GeV. Yet closer scrutiny cast serious doubts on such a scenario. The OPE treatment involves  applying quark-hadron duality. The latter can be 
expressed as saying that a rate evaluated on the quark-gluon level can be equated with the sum 
of observable exclusive hadronic channels, at least after some averaging or `smearing' over 
energy scales has been applied; the latter is sometimes referred to as semi-local duality. 

The question is: Which exclusive channels could be seen as dual to contributions with hadronic mass below 2.4 or even 2.5 GeV? (i) The theoretical estimates agree that the $D^{1/2}$ states that can populate this mass range possess too small production amplitudes to contribute significantly; their amplitudes actually 
would have to be enhanced greatly to overcome their kinematic suppression in semileptonic $B$ decays. (ii) There are many higher orbital excitations of course, and 
taken together they might have a `fighting' chance to yield a significant contribution to 
$\Gamma _{SL}(B)$ -- yet they all lie above 2.5 GeV in mass. It would not correspond to our usual picture  that exclusive channels all above 2.5 GeV are dual to quark-level contributions computed to all lie below it. (iii) There is one intriguing possibility: We know of one basic failure of all quark models: they cannot explain the mass of the baryonic Roper resonance. For all quark models predict the mass of the first radial excitation higher than that for the 
first orbital excitation -- an inequality clearly reversed for the Roper resonance. Would it be possible that the spectrum of charm resonances exhibits an analogous effect for mesons meaning that radial excitations can lie below the $P$ wave states discussed before, and they make up the $D_X$ component? This would be a most intriguing -- yet also most exotic explanation.  
\end{itemize}

{\em In summary:} ALEPH, DELPHI and D0 agree in finding a rate of about 0.8 - 1 \% of $\Gamma _B$ 
for the production of the two {\em narrow} $D^{**}$ states combined. This value is quite consistent 
with theoretical expectations for the $D^{3/2}$ rates. BELLE's data also fit naturally into this picture. 
The problem arises in the production of the {\em broad} $D^{**}$ states: The rate found by 
ALEPH and DELPHI suffice to saturate $\Gamma _{SL}(B)$, yet exceed the predictions for 
$\bar B \to l \bar \nu D^{1/2}$ by about an order of magnitude. BELLE's numbers on the other 
hand agree reasonably well with predictions, yet fall short of saturating $\Gamma _{SL}(B)$. 
The mass distributions of the broad $D^{**}$ states, 
for which there is no clear experimental verdict, might pose a theoretical conundrum: if it is centered 
below 2.5 GeV, we have no natural candidates for these states. 

\section{Comments on nonleptonic $B$ decays}
\label{NLPAT}

It is usually argued -- with very valid reasons -- that the theoretical description is much murkier for  nonleptonic than semileptonic $B$ decays. We might encounter here one of the few exception to this general rule of thumb, and we have been alluding to this possibility already. One can analyze the inclusive transition 
\beq 
\bar B \to \pi X_c
\eeq
and study the hadronic system $X_c$ in the spirit of factorization; i.e., one analyzes its recoil mass spectrum, its quantum numbers and decay characteristics with the following motivation: (i) The higher complexity of nonleptonic dynamics can be seen as an 
actually advantage here. For it provides additional production scenarios. More specifically it allows 
for significant production of $D^{1/2}$ states through the $W$ emission diagram, which is 
not possible in semileptonic transitions. (ii) Without a neutrino in the final state it might be easier 
to determine the mass distributions and quantum numbers of $X_c$. 

As already mentioned, for 
$X_c = D^*$, $D^{3/2,1/2}$ the theoretical predictions from the BT model  have been found in 
reasonable, though not compelling agreement with the data. 

One wants to extend such studies to the fully inclusive case and analyze also higher mass $D^{**}$ 
configuration. Intriguing first steps in this direction have already been undertaken by BABAR 
\cite{BABARNL,AMINA}. The recoil mass spectra obtained by BABAR show clear $D$ and $D^*$ peaks for charged and neutral $B$ decays. The former exhibit also a clear peak around 2.4 - 2.5 GeV and maybe a signal also in the region above 2.6 GeV. In 
the $\bar B_d$ case there might be a signal in the 2.4 - 2.6 GeV region, but not much else. 
The peaks in the 2.4 - 2.5 GeV region are natural candidates for showing 
$D^{3/2}$ production. The verdict on the domain beyond 2.5 GeV, which could be populated by the 
same configurations as in $\bar B \to l^-\bar \nu D^{**}_{broad}$, is tantalizing inconclusive.

\section{Conclusions and a call for action}
\label{ACTION}

The $B_d$ and $B_u$ inclusive semileptonic widths have been well measured. Most if not even all of it 
has been identified in $\bar B \to l \bar \nu D/D^* +(0,1)\pi$. The theoretical description of 
$\bar B \to l \bar \nu X_c$ rests on solid foundations. The potential discrepancies discussed in this note,  
which affect at most 20\% of semileptonic $B$ transitions, 
cannot lead to a significant increase in the uncertainty with which $|V(cb)|$ can be extracted from 
$\Gamma (\bar B \to l \bar \nu X_c)$. On the other hand they should not be `brushed under the rug'. 
Theory does make non-trivial predictions of a rather sturdy nature. The OPE treatment is genuinely based on QCD, and while the BT description invokes a model, it 
implements QCD dynamics for heavy flavour hadrons to a remarkable degree. Their predictions  therefore deserve to be taken seriously and not discarded at the first sign of phenomenological 
trouble. Preliminary lattice studies show no significant enhancement of 
`1/2' production. The numbers we have given for the theoretical expectations should be taken with 
quite a few grains of salt. Yet the predicted pattern that the abundance of `3/2' P wave resonances 
dominates over that for `1/2' states in semileptonic $B$ decays is a robust one. Even a failure of such well-grounded predictions could teach us valuable lessons on non-perturbative dynamics and our control over them; it would certainly provide a valuable challenge to lattice QCD.  Yet there is more: 
we know that $D$, $D^*$ and $D^{3/2}$ production do not saturate $\Gamma_{SL}(B)$, and theory 
tells us that $D^{1/2}$ cannot contribute significantly. What is then the nature of the missing hadronic 
configurations? 

On the experimental side the next important steps are:  
\begin{itemize}
\item 
In some of our discussion above we have modeled $\Gamma_{SL}(\bar B)$ as the incoherent sum of 
$\bar B \to l^-\bar \nu D/D^*/D^{3/2}/D_X$ to infer the average mass of the 
configuration $D_X$ and its variance from the measured hadronic mass moments. 
The values of $\langle M(D_X)\rangle$ and $\sqrt{\langle M^2(D_X)\rangle - \langle M(D_X)\rangle^2}$ 
serve as very useful diagnostics of the underlying dynamical situation. One finds that 
$\langle M(D_X)\rangle$ varies from 2.4 to 2.6 -- or even 2.7 -- GeV depending on whether one uses 
the branching ratios of Eq. (\ref{HFAGAVE}) or of Eq. (\ref{BDD*}). The main reason for this relatively 
sizable shift is the variation in BR$(\bar B \to l^-\bar \nu D^*)$. It would be most helpful to have this branching ratio determined more precisely. 
\item 
Clarifying the size, mass distribution and quantum numbers of 
$\bar B \to l \bar \nu [D/D^*\pi]_{\rm broad}$ and searching for 
$\bar B \to l \bar \nu D/D^* + 2 \pi$ with even higher sensitivity. 
\item 
The data should be presented {\em separately} for $\bar B\to l \bar \nu D+\pi$'s and 
$\bar B\to l \bar \nu D^*+\pi$'s, 
since it provides more theoretical diagnostics. 
\item 
More detailed analysis of $\bar B \to \pi X_c$, in particular in the high mass region for the $X_c$ 
and separately for charged and neutral $B$ decays \cite{BABARNL,AMINA}.

\end{itemize}
These are challenging experimental tasks, yet highly rewarding ones as well: 
\begin{itemize}
\item 
They probe our theoretical control over QCD's nonperturbative dynamics in novel and sensitive ways. 
This is an area where different theoretical technologies -- the OPE, quark models and 
lattice QCD -- are making closer and closer contact. 

The lessons to be learnt will be very significant ones, no matter what the eventual experimental verdict will be: 
\begin{itemize}
\item 
A confirmation of the OPE expectations and even the more specific BT predictions would reveal an even higher degree of theoretical control over nonperturbative QCD dynamics than has been 
shown through $\Gamma (\bar B \to l \bar \nu X_c)$. 
\item 
Otherwise we could infer that formally nonleading $1/m_Q$ corrections are highly significant numerically. Those corrections had to be highly enhanced to overcome the kinematic suppression in the 
production of $D^{1/2}$ in semileptonic $B$ decays. 

Such an insight would be surprising -- yet important as well. In particular 
it would provide a highly nontrivial challenge to lattice QCD. Meeting this challenge 
successfully would provide 
lattice QCD with significantly enhanced validation. 
\end{itemize} 
\end{itemize}
The call for further action is directed to theorists as well: 
\begin{itemize}
\item 
In the BT model one can compute the production rates for the {\em higher} orbital and radial 
excitations in semileptonic $B$ decays. The individual rates seem to be rather small. It is conceivable 
that summing over a multitude of them yields a significant contribution. 

\item 
The BT model predictions were obtained in the heavy quark limit. Corrections to this limit 
could be quite important as suggested 
in Ref.\cite{lig}, and they could significantly change the relative weight of 
$\tau ^{(n)}_{1/2}$ and $\tau ^{(n)}_{3/2}$. Calculating or at least constraining those corrections  would be a most worthwhile undertaking -- 
alas it requires some new ideas. A priori one can conceive of different ways of extending the 
BT description to include finite mass effects, yet they are unlikely to be equivalent. The foundations 
for a promising way have been laid in Ref.\cite{OPTICAL}.

The authors of Refs. \cite{EBERT} find significant rates for the production of radial excitation that are 
enhanced further by $1/m_Q$ corrections. While we cannot quite follow their argumentation, 
$1/m_Q$ corrections due to potential exchange diagrams could indeed be sizable. This intriguing 
possibility needs and deserves intense scrutiny. 

There is also the exotic possibility that one finds here a mesonic analogue to the Roper resonance, namely that the radial (and even other) excitations of charm mesons are significantly lower in mass than predicted by quark models. This would presumably raise their production rates 
in semileptonic $B$ decays to the sought-after level. 
\item 
Lattice QCD studies of `1/2' and `3/2' production in semileptonic $B$ decays has to be 
pursued with vigour. Such studies could turn out to be veritable `gold mines' as far as validation is concerned.  One can evaluate the spectrum of the higher radial and orbital excitations $D^{**}$, for which some encouraging results have already been obtained \cite{UKQCD-Trinlat}. Lattice 
calculations at finite values of $m_c$ should be performed, which would teach us about 
$1/m_c$ corrections. 

\item 
The strong decays $D^{**} \to D/D^* + \pi \pi$ should be estimated using heavy quark symmetry 
arguments augmented by quark model considerations. 

\end{itemize}

A final comment: The experimental analyses we advocate require considerable effort. We strongly belief 
such an effort is mandated by the insights to be gained, even if they are of a subtle nature: 
\begin{itemize}
\item 
They can provide us with important insights into the workings of nonperturbative dynamics. Lessons on the 
significance of $1/m_Q$ corrections and on systematic short comings of quark models would be of general value for the theoretical control we can establish over heavy flavour dynamics.
\item 
They probe the subtle concept of quark-hadron duality in novel ways. 
\item 
The greatest practical gain might emerge for lattice QCD: If the latter can meet the challenge of such detailed data successfully, it would have gained a qualitatively new measure of validation. 
\end{itemize}

\vspace{0.5cm}

\noindent
{\bf Acknowledgments:} This work was supported  by the NSF under grant number PHY-0355098 
and by the EC contract HPRN-CT-2002-00311 (EURIDICE).


\end{document}